\documentclass[12pt]{extarticle}%
\usepackage[a4paper, margin=1in]{geometry}
\usepackage{eurosym}
\usepackage{setspace}
\usepackage[utf8]{inputenc}
\usepackage{natbib}
\usepackage{amsmath,amsfonts,amssymb,amsthm,enumitem}
\usepackage{graphicx}
\usepackage{bibunits}
\usepackage{rotating}
\usepackage{subcaption}
\usepackage{multirow}
\usepackage{amsthm}
\usepackage{pgf}
\usepackage{setspace}
\usepackage{hyperref}
\usepackage{bbm}
\usepackage{comment}
\usepackage{tikz,calc}
\usepackage{accents}
\usepackage{xcolor}
\usepackage{soul}
\usepackage{pgfplots}
\usepackage{graphicx}
\usepackage{caption}
\usepackage{pdflscape}
\usepackage{float}
\usepackage{array}
\usepackage{multirow}
\usepackage{needspace}
\usepackage{placeins}
\usepackage{amsmath}
\usepackage{amsfonts}
\usepackage{xcolor}
\usepackage{amssymb}%
\providecommand{\U}[1]{\protect\rule{.1in}{.1in}}

\newtheorem{proposition}{Proposition}

\usepackage{authblk}
\usepackage{booktabs}
\usepackage{siunitx}
\usepackage{float}   
\usepackage{array}   
\usepackage{epigraph}
\newcolumntype{T}[1]{S[table-format=#1,table-number-alignment=center]@{}l@{}}

\begin{document}

\title{Chasing Opportunity: Spillovers and Drivers of U.S. State Population Growth}

\author[1]{Sebastian Kripfganz}
\author[2]{Vasilis Sarafidis\thanks{Corresponding author: Vasilis Sarafidis, Brunel University of London, email: vasilis.sarafidis@brunel.ac.uk}}

\affil[1]{University of Exeter}
\affil[2]{Brunel University of London}

\date{January 2025}
\maketitle

\begin{abstract}
We study the drivers and spatial diffusion of U.S. state population growth using a dynamic spatial model for 49 states, 1965–2017. Methodologically, we recover the spatial network structure from the data, rather than imposing it a priori via contiguity or distance, and combine this with an IV estimator that permits heterogeneous slopes and interactive fixed effects. This unified design delivers consistent estimation and inference in a flexible spatial panel model with endogenous regressors, a data-inferred network structure, and pervasive cross-state dependence. To our knowledge, it is the first estimation framework in spatial econometrics to combine all three elements within a single setting. 
Empirically, population growth exhibits broad yet heterogeneous conditional convergence: about three-quarters of states converge, while a small high-growth group mildly diverges. Effects of the core drivers, amenities, labour income, migration frictions, are stable across various network specifications. On the other hand, the productivity effect emerges only when the network is estimated from the data. Spatial spillovers are sizable, with indirect effects roughly one-third of total impacts, and diffusion extending beyond contiguous neighbours.
\end{abstract}



\baselineskip 0.27in

\newpage

\section{Introduction}
\begin{quote}\itshape
“Migration is the oldest action against poverty.” — Amartya Sen
\end{quote}

Population dynamics lie at the heart of regional development and long-term economic planning. Understanding what drives population growth and how it diffuses across space is essential for designing effective infrastructure policies, housing strategies, and public service provision. Rapid population growth can create pressures on roads, schools, utilities, and healthcare systems, while stagnation or decline can lead to underutilised infrastructure and economic contraction. Moreover, spatial shifts in population have direct implications for environmental sustainability, land use, and food security, particularly as population pressures interact with limited natural resources. In the United States, large and persistent differences in population growth across states have long reflected variation in economic opportunities, quality of life, and migration patterns. Uncovering the drivers of these differences and the extent to which growth in one region spills over to others is therefore central to understanding the evolving spatial structure of the U.S. economy.

Classic surveys in the population literature, such as \citet{Greenwood1997}, document that migration has historically served as an equilibrating mechanism, reallocating people toward locations offering higher expected returns net of moving costs. 
Urban studies have subsequently emphasized the role of consumption amenities and quality of life in shaping residential location decisions \citep[e.g.,][]{Roback1982,GlaeserKolkoSaiz2001,AlbouyEtal2016}.
For the United States, empirical work documents climate and amenity pull factors and the rise of Sunbelt destinations \citep{Rappaport2007,GlaeserTobio2008}, alongside job opportunities and changing urban fortunes \citep{GlaeserShapiro2003,Diamond2016}. Complementary evidence points to the role of job mobility and labour market adjustments in shaping the magnitude and intensity of internal migration patterns within the U.S. \citep{MolloySmithWozniak2011,MolloySmithWozniak2017}. 

Building on the preceding evidence, it is clear that the evolution of regional populations reflects spatial diffusion processes, whereby demographic and economic shocks in one area propagate through networks of migration, trade, and market integration. Spatial econometric approaches provide a coherent empirical framework for analysing such interdependencies, linking outcomes in one region to conditions and dynamics in others \citep{Elhorst2014}.
Against this backdrop, recent advances in econometrics have sought to broaden spatial frameworks along three important dimensions. 

\noindent
The first concerns heterogeneity in slope parameters and marginal effects across spatial entities. As noted by \citet{LeSageChih2016}, ``\textit{space–time panel data samples covering longer time spans allow us to produce parameter estimates for all N spatial units, an exciting point of departure for future work. Allowing for heterogeneous coefficients for each spatial unit holds a natural appeal when contrasted with conventional static spatial panel models.}'' Recent econometric work by \citet{AquaroBaileyPesaran2021} and \citet{ChenShinZheng2022} shows that slope heterogeneity is empirically and inferentially important in panel data analysis of spatial interactions, even after controlling for non-local, pervasive cross-sectional dependence driven by latent common factors.

\noindent
The second dimension concerns the specification of the spatial weighting matrix, $\mathbf{W}$. Two considerations motivate moving beyond setting it a priori on the basis of pure geographic contiguity schemes: first, such schemes may, by themselves, fail to capture interaction patterns shaped by trade, migration, or infrastructure. Second, even where geographic contiguity is empirically relevant, the functional form relating distance to interaction strength is not always known; shared-border indicators, inverse-distance kernels, and threshold rules remain, at best, approximations \citep{ElhorstTziolasTanMilionis2024,TanKesinaElhorst2025}. Accordingly, estimating $\mathbf{W}$ from the data aligns the network with observed interactions, accommodates non-geographic channels, and mitigates potential misspecification.

\noindent
A third methodological development recognises that, beyond spatial interactions, regional entities (and spatial units more generally) are subject to pervasive cross-sectional dependence arising from economy-wide or sector-specific shocks \citep{SarafidisWansbeek2012,SarafidisWansbeek2021}. To accommodate these influences, a dominant strategy is the latent common-factor (``interactive fixed effects'') approach, which captures co-movements beyond what spatial proximity alone can explain \citep{Sarafidis2007,PesaranTosetti2011,Yang2021,BaltagiShu2025}. For instance, demographic, institutional, macroeconomic, or technological shocks can affect all regions on a wide scale, albeit with heterogeneous intensities. 

A growing body of research on regional population growth supports all three dimensions of methodological development. For example, \citet{ReiaEtal2022} report substantial spatial heterogeneity in population dynamics within and across metropolitan areas, while county- and state-level analyses reveal that amenity-driven and network-mediated migration flows often extend well beyond geographic contiguity \citep{PartridgeEtal2008,ChiMarcouiller2013}. 
Motivated by this evidence, the present paper develops a spatial panel data model of population growth across 49 U.S. states over a 52-year period, integrating several features essential to capturing the complex and interdependent nature of regional population dynamics. 

First, we allow for endogenous spatial spillovers in population growth, recognising that migration and economic linkages need not conform to geographic borders. Accordingly, we go beyond purely geography-based weighting schemes and estimate the spatial connectivity matrix directly from the data. This step is important because population mobility may reflect economic and social networks rather than mere physical adjacency among states. 

Second, we permit heterogeneous, state-specific coefficients on the economic drivers of population growth; namely, total factor productivity, amenities, labour income, and migration costs. This is motivated by structural differences across states in industrial composition, demographic profiles, labour market flexibility and housing supply elasticity \citep{BlanchardKatz1992,GlaeserGyourko2005,Saiz2010,HsiehMoretti2019}. 

Third, we incorporate interactive fixed effects to capture nationwide demographic and institutional shocks, such as changes in fertility and ageing patterns, shifts in federal immigration policy, or national housing and credit cycles, which influence population growth across all states but with heterogeneous intensities. 

To jointly accommodate (i) an endogenously estimated spatial connectivity matrix, (ii) state-specific slope heterogeneity, and (iii) time-varying common shocks with heterogeneous loadings, we combine the Boosting One-Link-at-a-Time with Multiple Testing (BOLMT) approach of \citet{JuodisKapetaniosSarafidis2025} with the spatial Mean Group IV (MGIV) approach of \citet{ChenCuiSarafidisYamagata2025}.
 This integrated framework yields consistent estimation and inference in dynamic spatial panels with endogenous regressors and latent common factors. Stata implementations are available in \citet{KripfganzSarafidis2021,KripfganzSarafidis2025}. To our knowledge, this is the first empirical framework in the spatial econometrics literature that jointly integrates all three elements within a single setting.

Our empirical results show that U.S. state population growth is characterised by broad yet heterogeneous conditional convergence: three-quarters of states show convergence, whereas a handful of high-growth states display mild divergence. Coefficients on the core economic drivers (amenities, labour income, and migration frictions) are stable across alternative network specifications. By contrast, the effect of total factor productivity is significant only when the spatial network is estimated from the data and not under \emph{a priori} networks. Allowing for slope heterogeneity (MGIV) materially amplifies effect sizes relative to pooled estimators, indicating that homogeneous estimators understate responsiveness to fundamentals (amenities, labour income, and migration frictions). Spatial spillovers are economically meaningful and indirect effects account for about one-third of total marginal impacts. The estimated interaction network is sparse ($\approx$0.66\% density) but nation-wide in scope, and displays pronounced regional homophily, with same-division links occurring more than twice as often as under random assignment. Finally, latent common factors account for at least 60\% of the residual variance, even after controlling for state-specific and common time effects, highlighting the importance of modelling both network dependence and pervasive cross-sectional comovement.

\textit{Algorithmic extensions introduced in this study.} As a further contribution of this paper, we extend \texttt{spxtivdfreg} along two dimensions. First, we allow the spatial weighting matrix to vary over time; i.e., $\mathbf{W}$ becomes $\mathbf{W}_{t}$ so that network linkages can evolve. Second, we add post-estimation routines for \emph{spill-in} and \emph{spill-out} effects, complementing the standard direct/indirect decomposition (see, e.g., \citet{LeSageChih2016}, \citet{LeSagePace2019}, \citet{KrauseKripfganz2025}, \citet{AmendolagineProtaSerlenga2024}). Both extensions retain full backward compatibility with existing syntax. Further implementation details are provided in \citet{KripfganzSarafidis2025}.

\section{Model Specification and Motivation}\label{model}

Our empirical analysis uses the annual panel compiled by \citet{KleinmanLiuRedding2023}, covering the 49 contiguous U.S. states plus the District of Columbia from 1965 to 2017, with population levels and bilateral trade and migration flows.

\subsection{Dynamic Spatial Panel Data Model with Common Factors}

We study the following model:
\begin{equation}
\Delta y_{i,t} \;=\; \delta_{i}\, y_{i,t-1} \;+\; \psi_{i}\!\sum_{j=1}^{N} w_{i,j}\,\Delta y_{j,t} \;+\; \sum_{l=1}^{k} \beta_{l,i}\, x_{l,i,t} \;+\; u_{i,t}\,,
\label{eq:empirical}
\end{equation}
where $i=1,\ldots,N(=49)$ and $t=1,\ldots,T(=52)$. The outcome $\Delta y_{i,t}=\ln(Pop_{i,t})-\ln(Pop_{i,t-1})$ is population growth; $y_{i,t-1}=\ln(Pop_{i,t-1})$ represents the log-level of population at the beginning of the current growth spell, capturing the state’s starting
point in its growth trajectory; and $\sum_{j} w_{i,j}\Delta y_{j,t}$ is the network-weighted average growth of states connected to $i$ in the spatial weighting matrix.

The covariates are
(i) $x_{1,i,t}=TFP_{i,t}$: total factor productivity inferred by \citet{KleinmanLiuRedding2023} by inverting their dynamic spatial general equilibrium (DSGE) model;\footnote{\citet{KleinmanLiuRedding2023} assume Cobb–Douglas with labour and capital; see their online supplement for the construction of productivity and other model-implied objects.}
(ii) $x_{2,i,t}=Amenities_{i,t}$: model-implied state amenities from observed trade and migration flows \citep{KleinmanLiuRedding2023};
(iii) $x_{3,i,t}=\ln(Income_{i,t})$: labour income (nominal compensation of employees, millions of USD) per million people, obtained from the \emph{U.S. Bureau of Economic Analysis};
and (iv) $x_{4,i,t}=\ln(MigCost_{i,t})$: inbound migration frictions obtained from the \citet{KleinmanLiuRedding2023} DSGE model, computed as an inbound-migration–weighted measure and assuming that frictions are symmetric.

The error term is assumed to obey
\begin{equation}
u_{i,t} = \boldsymbol{\lambda}_{i}^{\prime}\mathbf{f}_{t} + \boldsymbol{\varphi}_{i}^{\prime}\mathbf{g}_{t} + \varepsilon_{i,t},
\label{eq:empirical_u}
\end{equation}
with latent common factors $\mathbf{f}_{t},\mathbf{g}_{t}$ and heterogeneous loadings $\boldsymbol{\lambda}_{i},\boldsymbol{\varphi}_{i}$. We include these interactive fixed effect components to capture nationwide demographic and institutional shocks, such as changes in fertility and aging patterns, shifts in federal immigration policy, and national housing and credit cycles, which affect all states but with heterogeneous intensities. The strength of each state’s response depends on structural features such as labour-market flexibility and housing-supply elasticity, which are hard to observe consistently across states; accordingly, we treat factors and factor loadings as latent and estimate them from the model.\footnote{We distinguish between $\mathbf{f}_{t}$ and $\mathbf{g}_{t}$ solely for generality, allowing some latent factors governing $y_{i,t}$ and $\mathbf{x}_{i,t}=\left(x_{1,i,t},\dots,x_{k,i,t}\right)^{\prime}$ to differ. This distinction is explained in Section \ref{Sec:Estimation_Methodology}, under Eq. \eqref{eq:spm_X_scalar}. Additional state and time fixed effects can be treated either explicitly or absorbed by the factor structure; Section \ref{Sec:Results} documents that the former approach is used in the present paper.}

There are three primary sources of endogeneity in Eq. \eqref{eq:empirical}. 
First, the spatial lag term $\sum_{j=1}^N w_{i,j}\,\Delta y_{j,t}$ is endogenous by construction: states’ growth rates are jointly determined within the spatial network, so each unit’s outcome depends on its neighbours' outcomes, which in turn depend on the unit's own outcome \citep[a simultaneity/reflection problem; see][]{Elhorst2021}. 
Second, the lagged dependent variable $y_{i,t-1}$ is endogenous in the presence of state-specific effects, generating the well-known dynamic panel (Nickell-type) bias even under strictly exogenous regressors \citep[e.g.,][]{Nickell1981,PhillipsSull2007}. 
Third, covariates may be correlated with the latent common factors in Eq. \eqref{eq:empirical_u}, so failing to control for (or instrument against) these factors induces omitted-variable bias.

The individual-specific structural coefficients play distinct roles. 
The coefficient $\delta_{i}$ governs transitional dynamics and conditional convergence: population converges if $\delta_{i}<0$ for all $i$, and diverges if $\delta_{i}>0$. When $\delta_{i}=0$, growth rates fluctuate around long-run paths that remain distinct across states.
The vector $\boldsymbol{\beta}_{i}$ encapsulates the immediate effects/impacts of a state's own covariates before any feedback effects propagating through the states' network. $\psi_{i}$ measures the strength of contemporaneous spillovers transmitted through neighbours’ growth \citep[e.g.,][]{KelejianPiras2017,JingEtAl2018,ArbiaEtal2010,ArbiaEtal2021}.\footnote{Our main results extend naturally to models with ``contextual effects'' \citep[e.g.,][]{Manski1993}, also known as the spatial Durbin model \citep{Elhorst2014}. In the Online Appendix, we estimate specifications that allow for such effects, but find no evidence of statistically significant contextual spillovers.}

\subsection{Spatial Weighting Matrices}

The spatial weights $w_{i,j}$ capture the strength of the directional linkages from state $j$ to state $i$. We construct a range of spatial weighting matrices $\mathbf{W}$ with elements $w_{i,j}$ in the $i$th row and $j$th column. These include a conventional binary contiguity matrix $\mathbf{W}_{1}$, in which two states are treated as neighbours if they share a common border, as well as two inverse-distance matrices $\mathbf{W}_{2}$ and $\mathbf{W}_{3}$. The latter are derived from the great-circle distance between the capital cities of states $i$ and $j$, denoted by $d_{i,j}$, which is computed using the Haversine formula based on their geographic coordinates. In this case, the spatial weights for both $\mathbf{W}_{2}$ and $\mathbf{W}_{3}$ are defined as:
\begin{equation}\label{eq:w_ij}
w_{i,j} = 
\begin{cases}
1/d_{i,j}, & \text{if } d_{i,j} \leq c \text{ miles} \\
0,         & \text{otherwise}
\end{cases}
\end{equation}
where $c$ denotes the $c$th percentile of the empirical distribution of $d_{i,j}$ across all state pairs $(i,j)$. This formulation imposes a hard distance threshold, restricting interactions to geographically proximate neighbours, with the intensity of interaction decaying inversely with distance. Specifically, $\mathbf{W}_{2}$ employs $c$ equal to the 10th percentile of the distance distribution, whereas $\mathbf{W}_{3}$ uses the 5th percentile, corresponding to network densities of approximately 10\% and 5\%, respectively.

Because geographic weights may not sufficiently capture the economic links between states \citep[e.g.,][]{Bavaud1998,CorradoFingleton2012,KrauseKripfganz2025}, we also construct a spatial weighting matrix $\mathbf{W}_{4}$ from bilateral trade flows with weights $w_{i,j} = 1/d_{i,j}$, where $d_{i,j}$ represents the average value of trade inflows from state $j$ to state $i$ over all years observed in the panel. This approach follows \citet{ErturKoch2011} and \citet{HoWangYu2013}, among others. Given that trade volumes can fluctuate substantially over time, a further specification uses a time-varying matrix $\mathbf{W}_{5}$ based on $w_{i,j,t} = 1/d_{i,j,t}$, where $d_{i,j,t}$ denotes trade inflows in period $t$. The resulting network densities for $\mathbf{W}_{4}$ and $\mathbf{W}_{5}$ are approximately 51\% and 52\% (averaged over $t$), respectively.\footnote{While $\mathbf{W}_{5}$ varies over time, we suppress the $t$ subscript for notational convenience. This also applies to the spatial weights $w_{i,j}$ in subsequent notation.} The origin-destination trade flow data for these matrices are taken from \citet{KleinmanLiuRedding2023}.

As is standard in spatial econometrics, all matrices $\mathbf{W}_{1},\dots,\mathbf{W}_{5}$ are constructed with zero diagonals $(w_{i,i} = 0)$ and are row-standardised so that the weights from all other states to state $i$ sum to one: $\sum_{j}w_{i,j}=1$ for each $i$. 

In addition to considering several a priori specifications for $\mathbf{W}$ (e.g., geographic contiguity or trade-inflow weights), we go beyond this practice by estimating the network structure directly from the data. Specifically, we employ the Boosting One-Link-at-a-Time with Multiple Testing (BOLMT) approach of \citet{JuodisKapetaniosSarafidis2025} to recover an endogenous spatial weighting matrix $\widehat{\mathbf{W}}$.

To facilitate exposition, consider the following static panel model with a spatial lag:
\begin{equation}
y_{i,t} \;=\; \psi_{i}\!\sum_{j=1}^{N} w_{i,j}\, y_{j,t} + \mathbf{x}_{i,t}^{\prime} \boldsymbol{\beta}_{i} + \varepsilon_{i,t}.
\label{eq:simpler_model}
\end{equation}
where $w_{i,i} = 0$ For later use, write $\mu_{i,t}=\ \sum\nolimits_{j=1}^{N}\omega
_{i,j}y_{j,t}$ and $\boldsymbol{\mu }_{i}=\left( \mu _{i,1},\mu _{i,2},...,\mu_{i,T}\right) ^{\prime }$.

The BOLMT procedure constructs $\widehat{\mathbf{W}}$ by running, for each unit $i$ an instrumental variables (IV) regression of $y_{i,t}
$ on $\mathbf{x}_{i,t}$ and $y_{j,t}$, $j=1, 2, \dots, i-1, i+1, \dots, N$, \emph{taken one at a time} and instrumented by $\mathbf{x}_{j,t}$.\footnote{While the BOLMT procedure is described here in generic terms, the specific instrument sets used in the pairwise IV regressions are detailed in Section~\ref{Sec:Estimation_Methodology} (see Eq.~\eqref{Z_matrix_empirical}). These instruments coincide with those employed for estimation of the structural slope parameters, thereby maintaining a unified identification strategy across network recovery and model estimation.} 

Let $t_{i,j}$ denote the $t$-ratio in that $i$-specific IV regression corresponding to the coefficient of $y_{j,t}$. With $\mathbf{P}%
_{\boldsymbol{Z}}=\boldsymbol{Z}\left( \boldsymbol{Z}^{\prime }\boldsymbol{Z}%
\right) ^{-1}\boldsymbol{Z}^{\prime }$ and $\mathbf{M}_{%
\boldsymbol{Z}}=\boldsymbol{I}_{T}-\mathbf{P}_{\boldsymbol{Z}}$, define $\boldsymbol{y}_{i}=\left( y_{i,1},y_{i,2},...,y_{i,T}\right) ^{\prime}$, $\boldsymbol{\widehat{y}}_{j}=\mathbf{P}_{\mathbf{x}_{j}}\boldsymbol{y}_{j}$, $\mathbf{x}_{j}\mathbf{=}\left(
\mathbf{x}_{j,1},\mathbf{x}_{j,2},...,\mathbf{x}_{j,T}\right)
^{\prime }$, and $\widehat{\sigma}_{i}$ as the standard error of the IV regression for unit $i$. Then,
\begin{equation}
t_{i,j}=\frac{T^{-1/2}\boldsymbol{\widehat{y}}_{j}^{\prime }\mathbf{M}_{%
\mathbf{x}_{i}}\boldsymbol{y}_{i}}{\widehat{\sigma}_{i}\sqrt{T^{-1}%
\boldsymbol{\widehat{y}}_{j}^{\prime }\mathbf{M}_{\mathbf{x}_{i}}\boldsymbol{%
\widehat{y}}_{j}}}\text{,}
\label{eq:ti}
\end{equation}%
where $\boldsymbol{\varepsilon }_{i}=\left( \varepsilon_{1},\dot,\varepsilon _{T}\right) ^{\prime }$. 
The statistic $t_{i,j}$ is the workhorse for selecting links into row $i$ of $\widehat{\mathbf{W}}$. Intuitively, if unit \(j\) truly influences \(i\) (i.e., \(w_{i,j}\neq 0\)), the IV regression of \(y_{i,t}\) on \(y_{j,t}\) (controlling for \(\mathbf{x}_{i,t}\)) will tend to yield a large \(\lvert t_{i,j}\rvert\).
The BOLMT algorithm proceeds in a stepwise, multiple-testing fashion. For a fixed \(i\), compute all
\(\{t_{i,1}, t_{i,2}, \ldots, t_{i,i-1}, t_{i,i+1}, \dots, t_{i,N}\}\) with \(j\neq i\) and rank them by absolute magnitude. Select the unit \(j^{\star}\) associated with the largest \(\lvert t_{i,j}\rvert\) if and only if it exceeds a data-driven threshold that controls the false positive rate. Upon selection, augment the regressor set for equation \(i\) to include \(y_{j^{\star},t}\)
(with appropriate instrumentation), recompute the IV \(t\)-statistics for the remaining candidates conditional on the expanded regressor set, and iterate. The procedure stops for row \(i\) when no remaining \(\lvert t_{i,j}\rvert\) exceeds the threshold. Repeating this cycle across all \(i=1,\ldots,N\) yields a fully specified network. Again, once $\widehat{\mathbf{W}}$ is obtained, all weights are row-standardised, such that the elements in each row sum to one. 
One then proceeds to Mean-Group IV estimation of the unit-specific coefficients, as discussed in the next section.\footnote{Note that when a spatial unit is found to have no neighbours, this individual equation is excluded from estimation of the spatial lag parameter $\psi$.}

A formal treatment is provided by \citet{JuodisKapetaniosSarafidis2025}, who establish \emph{selection consistency}: as \(N,T\to\infty\), the procedure recovers the true network almost surely. They further show that the Mean-Group IV estimator is consistent and asymptotically normal, with an asymptotic variance identical to that obtained under known $\mathbf{W}$.

In our application, the BOLMT procedure yields a sparse yet connected network: the estimated density is approximately \(0.66\%\) of the \(N(N-1)\) potential directed links.

\section{Estimation Methodology}\label{Sec:Estimation_Methodology}

Following the methodology of \citet{Pesaran2006}, \citet{CuiNorkuteSarafidisYamagata2022} and others, we permit the model covariates to be correlated with the common factor component. Therefore, we complement Eq. \eqref{eq:empirical} by assuming 
\begin{equation}  \label{eq:spm_X_scalar}
\mathbf{x}_{i,t}=\boldsymbol{\Gamma}_{i}^{\prime }\mathbf{f}_t+\mathbf{v}_{i,t},
\end{equation}
where $\mathbf{x}_{i,t}=\left(x_{1, i,t},\dots,x_{k, i,t}\right)^{\prime}$, $\boldsymbol{\Gamma}_{i}$ represents an $r_x \times k$ matrix of factor loadings and $\mathbf{v}_{i,t}$ is a purely idiosyncratic error that is independent from $\varepsilon_{i,t}$ in Eq. \eqref{eq:empirical_u}. Importantly, for the sake of generality, we allow some of the latent factors governing $y_{i,t}$ and $\mathbf{x}_{i,t}$ to differ. This distinction justifies the inclusion of the additional term $\boldsymbol{\phi}_i^{\prime }\boldsymbol{g}_t$ in model \eqref{eq:empirical}.\footnote{Alternatively, one may assume that the additional term $\boldsymbol{\phi}_i^{\prime }\boldsymbol{g}_t$ enters the process for $\mathbf{x}_{i,t}$ rather than $u_{i,t}$. This case is conceptually simpler, since all factors directly affecting $y$ would then be spanned by $\mathbf{x}_{i,t}$.} Moreover, the loadings $\boldsymbol{\Gamma}_{i}$ are permitted to exhibit correlation with both $\boldsymbol{\lambda}_i$ and $\boldsymbol{\phi}_i$, further accommodating potential interdependencies that may arise in the components of the model. 

Stacking Eq. (\ref{eq:empirical}) and \eqref{eq:spm_X_scalar} over $t$ yields
\begin{equation}  \label{pm_vectori}
\begin{split}
\Delta \mathbf{y}_{i}&=\delta_i \mathbf{y}_{i,-1} + \psi_i \sum_{j=1}^{N} w_{i,j} \Delta \mathbf{y}_{j} +%
\mathbf{X}_{i}\boldsymbol{\beta}_i+\mathbf{F}\boldsymbol{\lambda}_i+\mathbf{G%
}\boldsymbol{\phi}_i+\boldsymbol{\varepsilon}_{i}, \\
\mathbf{X}_{i}&=\mathbf{F}\boldsymbol{\Gamma}_{i}+\mathbf{V}_{i},\ \ \ i=1,\ldots,N,
\end{split}%
\end{equation}
where  $\Delta \mathbf{y}_{i}=(\Delta y_{i,1},\ldots,\Delta y_{i,T})^{\prime }$, $\mathbf{y}%
_{i,-1}=(y_{i,0},\ldots,y_{i,T-1})^{\prime }$, $\Delta \mathbf{y}_{j}=(\Delta y_{j,1},\ldots,\Delta y_{j,T})^{\prime}$, $\mathbf{X}_{i}=(\mathbf{x}_{i,1},\cdots,%
\mathbf{x}_{i,T})^{\prime }$, $\mathbf{F}=(\mathbf{f}_1,\ldots,\mathbf{f}%
_T)^{\prime }$, $\mathbf{G}=(\boldsymbol{g}_1,\ldots,\boldsymbol{g}%
_T)^{\prime }$, $\boldsymbol{\varepsilon}_{i}=(\varepsilon_{i,1},\ldots,%
\varepsilon_{i,T})^{\prime }$ and $\mathbf{V}_{i}=(\mathbf{v}_{i,1},\ldots,%
\mathbf{v}_{i,T})^{\prime }$.\footnote{When spatial weights are time-varying, let $\{\mathbf W_t\}_{t=1}^T$ denote the sequence of $N\times N$ spatial weighting matrices, and define $\mathbf w_{i,j} \equiv (w_{i,j,1},\dots,w_{i,j,T})^{\prime}$ as the $T\times1$ vector collecting the $(i,j)$th element of $\mathbf W_t$ over time. The spatial lag of $\Delta \mathbf y_i$ is then given by $\sum_{j=1}^N \mathbf w_{i,j} \odot \Delta \mathbf y_j$, where $\odot$ denotes the Hadamard (element-wise) product. An analogous construction applies to spatial lags of $\mathbf X_i$ used as instruments in Eq.~(\ref{Z_matrix_empirical}). No further modification of the estimation procedure is required. To avoid unnecessary notational complexity, we present the exposition under time-invariant spatial weights.}

Define $\boldsymbol{\theta}
_i=(\delta_i,\psi_i,\boldsymbol{\beta}_i^{\prime })^{\prime }$, and $\mathbf{C}%
_{i}=(\mathbf{y}_{i,-1},\sum_{j=1}^{N} w_{i,j}  \Delta \mathbf{y}_{j},\mathbf{X}_{i})$. Then, the stacked model in Eq. (\ref{pm_vectori}) can be reformulated as
\begin{equation}
\mathbf{y}_{i}=\mathbf{C}_{i}\boldsymbol{\theta}_i+\mathbf{u}_{i}.
\end{equation}
where $\mathbf{u}_{i}=\mathbf{F}\boldsymbol{\lambda}_i+\mathbf{G}\boldsymbol{%
\phi}_i+\boldsymbol{\varepsilon}_{i}$.

We use an Instrumental Variables (IV) method to estimate $\boldsymbol{\theta}_{i}$. The individual-specific IV estimator of $\boldsymbol{\theta}_i$ is given by
\begin{equation}\label{individual-est}
\widehat{\boldsymbol{\theta}}_i=\left(\widehat{\mathbf{A}}_i^{\prime }%
\widehat{\mathbf{B}}^{-1}_i\widehat{\mathbf{A}}_i\right)^{-1}\widehat{%
\mathbf{A}}_i^{\prime }\widehat{\mathbf{B}}_i^{-1}\widehat{\mathbf{c}}_{i},
\end{equation}
where
\begin{equation}\label{notation1}
\widehat{\mathbf{A}}_i=T^{-1}\widehat{\mathbf{Z}}_{i}^{\prime }
\mathbf{C}_{i},\ \ \ \ \widehat{\mathbf{B}}_i=T^{-1} \widehat{\mathbf{Z}}%
_{i}^{\prime }\widehat{\mathbf{Z}}_{i},\ \ \ \ \widehat{\mathbf{c}}_{i}=T^{-1}
\widehat{\mathbf{Z}}_{i}^{\prime }\mathbf{y}_{i},
\end{equation}
and
\begin{align}
\mathbf{\widehat{Z}}_{i} = \Big(
& {\mathbf{M}_{\widehat{\mathbf{F}}}}\mathbf{X}_{i}, \; \;
\mathbf{M}_{\widehat{\mathbf{F}}} \mathbf{M}_{\widehat{\mathbf{F}}_{-1}}\mathbf{X}_{i,-1}, \; \;
\mathbf{M}_{\widehat{\mathbf{F}}} \mathbf{M}_{\widehat{\mathbf{F}}_{-2}}\mathbf{X}_{i,-2}, \nonumber \\
& \mathbf{M}_{\widehat{\mathbf{F}}} \sum_{j=1}^N w_{i,j}\mathbf{X}_{j}, \; \;
\mathbf{M}_{\widehat{\mathbf{F}}} \mathbf{M}_{\widehat{\mathbf{F}}_{-1}}\sum_{j=1}^N w_{i,j}\mathbf{X}_{j,-1}, \; \;
\mathbf{M}_{\widehat{\mathbf{F}}} \mathbf{M}_{\widehat{\mathbf{F}}_{-2}}\sum_{j=1}^N w_{i,j}\mathbf{X}_{j,-2}
\Big),
\label{Z_matrix_empirical}
\end{align}
with $\mathbf{M}_{{\widehat{\mathbf{F}}_{-\tau}}}=\mathbf{I}_T - \widehat{\mathbf{F}}_{-\tau}(\widehat{\mathbf{F}}_{-\tau}^{\prime }\widehat{\mathbf{F}}_{-\tau})^{-1}\widehat{\mathbf{F}}_{-\tau}^\prime$ for $\tau=0,1,2$.\footnote{We use the convention that $\mathbf{F}_{-0}=\mathbf{F}$.} $\widehat{\mathbf{F}}_{-\tau}$ is obtained as $\sqrt{T}$ times the eigenvectors corresponding to the $r_x$ largest eigenvalues of the $T \times T$ matrices $(NT)^{-1}\sum_{i=1}^N \mathbf{X}_{i,-\tau}\mathbf{X}_{i,-\tau}^{\prime }$. 

Loosely speaking, $\mathbf{M}_{\widehat{\mathbf{{F}}}} \mathbf{M}_{\widehat{\mathbf{{F}}}_{-1}}\mathbf{X}_{i,-1}$ and $\mathbf{M}_{\widehat{\mathbf{{F}}}} \mathbf{M}_{\widehat{\mathbf{{F}}}_{-2}}\mathbf{X}_{i,-2}$ instrument $\mathbf{y}_{i,-1}$ in Eq. \eqref{pm_vectori}, ${\mathbf{M}_{\widehat{\mathbf{{F}}}}}\mathbf{X}_{i}$ instruments $\mathbf{X}_{i}$, and the remaining terms instrument the spatial lag. Since $k=4$, this specification yields a total of 24 instruments.

Assuming that the cross-sectionally heterogeneous coefficients $\boldsymbol{\theta}_i$ follow the random-coefficient model, it is known that the dynamic pooled estimator of the population mean $\boldsymbol{\theta}=\mathbb{E}(\boldsymbol{\theta}_i)$ will be inconsistent \citep[see][]{RobertsonSymons1992,PesaranSmith1995}.\footnote{Within the spatial panel data literature, slope parameter heterogeneity is formally motivated and treated under a random coefficients model by \citet{ChenShinZheng2022}. While their framework is static and thus not directly applicable to the dynamic setting considered here, the same considerations regarding the inconsistency of pooled estimators and the interpretation of average effects under heterogeneity remain relevant.} In the present case, the same holds true even in the absence of temporal dynamics in model \eqref{pm_vectori} because the spatial lag variable, $\sum_{j=1}^{N} w_{i,j} \Delta \mathbf{y}_{j}$, is endogenous by construction. For this reason, we employ the MGIV approach developed by \citet{ChenCuiSarafidisYamagata2025}, which combines the individual-specific IV estimates and averages them to obtain consistent estimates of population-level average effects.

Specifically, once all $\widehat{\boldsymbol\theta}_i$, $1\leq i\leq N$,  are obtained, the MGIV estimator of $\boldsymbol{\theta}$ is computed as
\begin{equation}\label{eq-MGest}
\widehat{\boldsymbol\theta}_{MGIV}=\frac{1}{N}\sum\limits_{i=1}^N\widehat{\boldsymbol\theta}_i.
\end{equation}

Proposition \ref{MG-asydist} below establishes the asymptotic properties of $\widehat{\boldsymbol\theta}_{MGIV}$.

\begin{proposition}\label{MG-asydist}
Under certain regularity conditions listed in \citet{ChenCuiSarafidisYamagata2025}, as $N,T\rightarrow\infty$ with $N/T^2\rightarrow0$, the MGIV estimator, $\widehat{\boldsymbol\theta}_{MGIV}$, is consistent for the population mean $\boldsymbol{\theta}$. If, it further holds that $N/T^{6/5}\rightarrow0$, then the MGIV estimator has the following asymptotic distribution
\begin{equation*}
\sqrt{N}(\widehat{\boldsymbol\theta}_{MGIV}-\boldsymbol{\theta})\stackrel{d}{\longrightarrow} N\left(\mathbf{0},\ \boldsymbol{\Sigma}_{\boldsymbol\theta}\right),\ \ {\rm as\ }N,T\rightarrow\infty,
\end{equation*}
where $\boldsymbol{\Sigma}_{\boldsymbol\theta}$ denotes the variance-covariance matrix of $\widehat{\boldsymbol\theta}_{MGIV}$. Furthermore, 
\begin{align}
\widehat{\boldsymbol{\Sigma}}_{\boldsymbol\theta}=\frac{1}{N-1}\sum\limits_{i=1}^N \left(\widehat{\boldsymbol{\theta}}_i-\widehat{\boldsymbol\theta}_{MGIV}\right)\left(\widehat{\boldsymbol{\theta}}_i-\widehat{\boldsymbol\theta}_{MGIV}\right)^\prime \notag
\end{align}
is a consistent estimator of $\boldsymbol{\Sigma}_{\boldsymbol\theta}$.
\end{proposition}

In regression models with spatial lags, the coefficients $\beta_{l,i}$, $l=1,\ldots,k$, and their MG analogues only measure the immediate impact of the respective covariates, prior to any accumulation of feedback effects across states. Provided that $\delta \neq 0$, the reduced-form marginal effects are heterogeneous across states even when they are evaluated at the population mean $\boldsymbol{\theta}$ of the heterogeneous coefficients $\boldsymbol{\theta}_i$. Stacking all variables over $i$ for a given time period $t$ yields
\begin{equation}
\Delta \mathbf{y}_{(t)} =\delta \mathbf{y}_{(t-1)} + \psi \mathbf{W} \Delta \mathbf{y}_{(t)} +\mathbf{X}_{(t)}\boldsymbol{\beta}+\mathbf{u}_{(t)},%
\end{equation}
where $\Delta \mathbf{y}_{(t)}=(\Delta y_{1,t},\ldots,\Delta y_{N,t})^{\prime }$, $\mathbf{y}%
_{(t-1)}=(y_{1,t-1},\ldots,y_{N,t-1})^{\prime }$, $\mathbf{X}_{(t)}=(\mathbf{x}_{1,t},\ldots,%
\mathbf{x}_{N,t})^{\prime }$, and $\mathbf{u}_{(t)}=(u_{1,t},\ldots,u_{N,t})^{\prime }$. 

Under standard regularity conditions on the invertibility of matrix $\mathbf{S} (\psi) = \mathbf{I}_N - \psi \mathbf{W}$, which requires $|\psi| < 1$ in the case of a row-standardised spatial weighting matrix, the reduced-form representation of the model is
\begin{equation}
\Delta \mathbf{y}_{(t)}= \mathbf{S}^{-1} (\psi) \left( \delta \mathbf{y}_{(t-1)} + \mathbf{X}_{(t)}\boldsymbol{\beta}+\mathbf{u}_{(t)} \right).%
\end{equation}
Therefore, the marginal effects of the $l$th regressor are governed by $\beta_l \mathbf{S}^{-1} (\psi)$, and not just $\beta_l$. In the spatial econometrics literature, the main-diagonal elements of this matrix are referred to as direct (or own) effects, while the off-diagonal elements constitute indirect (or spillover) effects. It is common practice to report averages of those state-specific direct or indirect effects. The sum of the two then yields average total effects. However, especially when there are  strong indirect effects as a consequence of a significant spatial lag coefficient $\psi$, these average effects can mask substantial heterogeneity of the marginal effects across states. Firstly, depending on the network linkages, population growth in different states may respond with varying degrees to changes in the covariates, even if the same change occurs uniformly across all states. Secondly, the responsiveness of a state's population growth varies depending on the location where a change of the covariates occurs, if the latter is not uniform.

Following \citet{KrauseKripfganz2025}, we also compute individual responses of population growth in state $i$ to a change of a covariate in state $j$ by selecting the respective element in the $i$th row and $j$th column of the spatial multiplier matrix $\mathbf{S}^{-1} (\psi)$, multiplied by $\beta_l$. Averages can be formed over selected rows (or columns) of $\mathbf{S}^{-1} (\psi)$ to compute the impact on (or from) specific groups, and the heterogeneity of individual responses can be highlighted graphically. For example, it can be of interest to analyse the spatial distribution of states' responses to a ``shock'' occurring in the Midwest region of the U.S., rather than an equal shock originating everywhere simultaneously.


\section{Results}\label{Sec:Results}

We employ several estimators that differ in their maintained assumptions regarding slope homogeneity and the specification of the spatial weighting matrix. In all cases, we apply a two–way within transformation (deviations from unit and time means) to eliminate state fixed effects and common time effects. Consequently, the error term in Eq.~\eqref{eq:empirical_u} retains factors and factor loadings measured relative to their means.

The analysis has been conducted in Stata using \texttt{spxtivdfreg} \citep{KripfganzSarafidis2025}, which is an extension of \texttt{xtivdfreg} \citep{KripfganzSarafidis2021} to spatial settings. The command implements the IV framework of \citet{CuiSarafidisYamagata2023} and \citet{ChenCuiSarafidisYamagata2025} for spatial dynamic panel models with interactive effects and, where appropriate, heterogeneous slope coefficients. More generally, the command is designed to accommodate a broad class of spatial-dynamic specifications by allowing for additional temporal lags, spatial lags, alternative instrument constructions, and includes built-in routines for direct, indirect, and total spatial impacts. See \citet{KripfganzSarafidis2025} for further details.

\subsection{Homogeneous vs Heterogeneous Models}

Tables~\ref{tab:2siv} and~\ref{tab:mgiv} report estimation results based on the homogeneous spatial 2SIV estimator of \citet{CuiSarafidisYamagata2023} and the heterogeneous spatial Mean Group IV (MGIV) estimator of \citet{ChenCuiSarafidisYamagata2025}, respectively. Both approaches employ the instrument set $\mathbf{\widehat{Z}}_{i}$ and alternative spatial weighting matrices $\widehat{\mathbf{W}}$ and $\mathbf{W}_{1},\dots,\mathbf{W}_{5}$. In the homogeneous specification, it is assumed that $\delta_{i}=\delta$, $\psi_{i}=\psi$ and $\boldsymbol{\beta}_{i}=\boldsymbol{\beta}$.

The pooled (homogeneous) estimates in Table~\ref{tab:2siv} are remarkably stable across weighting schemes in terms of sign, magnitude, and statistical significance, suggesting that the findings are robust across the network specifications evaluated in this study. The only departure concerns the coefficient on total factor productivity (TFP), for which statistical significance depends on the selected weighting scheme. 
Moreover, differences in the spatial-lag coefficient across network specifications are natural, since the coefficient captures spillovers relative to the particular network employed, and hence depends on its scaling.
As a result, cross-scheme comparisons of these coefficients are not meaningful. Comparability is instead achieved through the model-implied effects (direct, indirect, and total), which we report in the next section; these effects aggregate the propagation mechanism under the relevant network structure and provide a common basis for interpretation.

The dimensionality of the factor structure in the first-stage IV residuals and the covariates, denoted by $r_y$ and $r_x$, respectively, is determined from the data using the eigenvalue method of \citet{AhnHorenstein2013}.
In all cases, two factors are selected for the covariates ($r_x = 2$) and one for the first-stage IV residuals ($r_y = 1$).
The estimated value of $\rho$, which measures the proportion of the variance of the total regression error attributable to the common factor component, exceeds 0.60 (60\%) in every specification. This finding suggests that even after controlling for both state-specific and common time effects, a substantial share of residual dependence remains, implying that the traditional two-way fixed-effects structure is insufficient to capture the full extent of cross-sectional dependence in the data.

Although the coefficient estimates are stable across specifications, the J-test rejects the null hypothesis of valid overidentifying restrictions in every case. This outcome is consistent with slope parameter heterogeneity across states, implying that the homogeneity assumption underlying the pooled estimator is too restrictive.

Table~\ref{tab:mgiv} presents results from the heterogeneous MGIV estimator, which explicitly allows for variation in slope parameters. Several noteworthy differences emerge when comparing these results with those from the homogeneous model.
First, the coefficient on lagged population ($y_{i,t-1}$) is notably larger (nearly three times as large in magnitude), implying that homogeneous estimators tend to underestimate the average strength of population convergence dynamics.

Second, the TFP coefficient is statistically significant only when the data-driven spatial weighting matrix $\widehat{\mathbf{W}}$ is used; it is insignificant under all alternative matrices. A similar pattern emerges for magnitudes of other coefficients: estimates are generally larger under MGIV than under 2SIV, and larger with $\widehat{\mathbf{W}}$ than with the \emph{a priori} W's, while statistical significance is otherwise broadly unchanged across specifications.
All estimators in Table~\ref{tab:mgiv} are strictly more general than the pooled 2SIV estimators in Table~\ref{tab:2siv}. Within this set, the $\widehat{\mathbf{W}}$ column, corresponding to MGIV with the network estimated from the data, offers the greatest flexibility relative to \emph{a priori} W's and is thus preferred empirically.

\begin{table}[H]
\centering
\caption{Homogeneous IV Estimation Results (2SIV)}
\label{tab:2siv}
\small
\begin{tabular}{lcccccc}
\toprule
 & \multicolumn{6}{c}{2SIV} \\
\cmidrule(lr){2-7}
 & $\widehat{\mathbf{W}}$ & $\mathbf{W}_{1}$ & $\mathbf{W}_{2}$ & $\mathbf{W}_{3}$ & $\mathbf{W}_{4}$ & $\mathbf{W}_{5}$ \\
\midrule
$y_{i,t-1}$ {\footnotesize (lagged population)} & -0.016*** & -0.022*** & -0.014*** & -0.019*** & -0.025*** & -0.025*** \\
         & (0.006)   & (0.005)   & (0.004)   & (0.005)   & (0.005)   & (0.005)   \\[3pt]
$x_{1,i,t}$ {\footnotesize (TFP)}     & 0.000   & 0.002     & 0.002**   & 0.003**   & 0.002**   & 0.001     \\
 & (0.001)   & (0.001)   & (0.001)   & (0.001)   & (0.001)   & (0.001)   \\[3pt]
$x_{2,i,t}$ {\footnotesize (amenities)} & 0.005***  & 0.006***  & 0.006***  & 0.006***  & 0.007***  & 0.006***  \\
        & (0.000)   & (0.001)   & (0.000)   & (0.000)   & (0.000)   & (0.000)   \\[3pt]
$x_{3,i,t}$ {\footnotesize (labour income)} & 0.038*** & 0.030*** & 0.033*** & 0.031*** & 0.032*** & 0.033*** \\
           & (0.004) & (0.004) & (0.005) & (0.005) & (0.003) & (0.004) \\[3pt]
$x_{4,i,t}$ {\footnotesize (migration cost)} & -0.002*** & -0.002*** & -0.002*** & -0.002*** & -0.002*** & -0.002*** \\
            & (0.000)   & (0.000)   & (0.000)   & (0.000)   & (0.000)   & (0.000)   \\[6pt]
$\sum_{j=1}^{N} w_{i,j} \Delta y_{j,t}$ {\footnotesize (spatial lag)} & 0.335*** & 0.083*** & 0.169*** & 0.135*** & 0.230* & 0.538*** \\
          & (0.038)  & (0.013)  & (0.037)  & (0.034)  & (0.125)  & (0.141)  \\[6pt]
\midrule
N & 49 & 49 & 49 & 49 & 49 & 49 \\
$\chi^2_J$ & 36.504 & 35.064 & 33.895 & 33.261 & 34.584 & 34.678 \\
$p_J$ & 0.006 & 0.009 & 0.013 & 0.016 & 0.011 & 0.010 \\
r$_x$ & 2 & 2 & 2 & 2 & 2 & 2 \\
r$_y$ & 1 & 1 & 1 & 1 & 1 & 1 \\
$\rho$ & 0.614 & 0.702 & 0.745 & 0.744 & 0.739 & 0.700 \\
\bottomrule
\end{tabular}
\vspace{0.5em}
\parbox{\linewidth}{
\small
\textit{\textbf{Notes}:} Standard errors in parentheses. * p$<$0.10, ** p$<$0.05, *** p$<$0.01. $\widehat{\mathbf{W}}$ denotes the data-driven estimated spatial network matrix obtained using the BOLMT approach of \citet{JuodisKapetaniosSarafidis2025}. $\mathbf{W}_{1}$ denotes the contiguity matrix; $\mathbf{W}_{2}$ and $\mathbf{W}_{3}$ denote inverse-distance matrices with 10th and 5th percentile cutoffs, respectively; $\mathbf{W}_{4}$ and $\mathbf{W}_{5}$ denote trade-based weighting matrices, with $\mathbf{W}_{4}$ time-averaged and $\mathbf{W}_{5}$ time-varying.
}
\end{table}

\begin{table}[H]
\centering
\caption{Heterogeneous IV Estimation Results (Mean Group)}
\label{tab:mgiv}
\small
\begin{tabular}{lcccccc}
\toprule
 & \multicolumn{6}{c}{MGIV} \\
\cmidrule(lr){2-7}
 & $\widehat{\mathbf{W}}$ & $\mathbf{W}_{1}$ & $\mathbf{W}_{2}$ & $\mathbf{W}_{3}$ & $\mathbf{W}_{4}$ & $\mathbf{W}_{5}$ \\
\midrule
$y_{i,t-1}$ {\footnotesize (lagged population)} & -0.051*** & -0.043*** & -0.043*** & -0.045*** & -0.047*** & -0.051*** \\
        & (0.015)   & (0.012)   & (0.012)   & (0.013)   & (0.013)   & (0.014)   \\[3pt]
$x_{1,i,t}$ {\footnotesize (TFP)}     & 0.005**    & 0.001     & 0.002     & 0.002     & 0.005     & 0.003     \\
        & (0.003)   & (0.003)   & (0.003)   & (0.003)   & (0.003)   & (0.004)   \\[3pt]
$x_{2,i,t}$ {\footnotesize (amenities)} & 0.007***  & 0.005***  & 0.005***  & 0.006***  & 0.006***  & 0.006***  \\
        & (0.001)   & (0.001)   & (0.001)   & (0.001)   & (0.001)   & (0.001)   \\[3pt]
$x_{3,i,t}$ {\footnotesize (labour income)} & 0.050*** & 0.047*** & 0.047*** & 0.048*** & 0.049*** & 0.050*** \\
           & (0.012)  & (0.010)  & (0.011)  & (0.011)  & (0.011)  & (0.010)  \\[3pt]
$x_{4,i,t}$ {\footnotesize (migration cost)} & -0.004*** & -0.002*** & -0.002*** & -0.002*** & -0.003*** & -0.003*** \\
            & (0.001)  & (0.001)  & (0.001)  & (0.001)  & (0.001)  & (0.001)  \\[6pt]
$\sum_{j=1}^{N} w_{i,j} \Delta y_{j,t}$ {\footnotesize (spatial lag)} & 0.357*** & 0.114*** & 0.509*** & 0.440*** & 0.733*** & 0.602** \\
          & (0.055)  & (0.014)  & (0.063)  & (0.042)  & (0.212)  & (0.295)  \\[6pt]
\midrule
N & 49 & 49 & 49 & 49 & 49 & 49 \\
r$_x$ & 2 & 2 & 2 & 2 & 2 & 2 \\
\bottomrule
\end{tabular}
\vspace{0.5em}
\parbox{\linewidth}{
\small
\textit{\textbf{Notes}:} Standard errors in parentheses. * p$<$0.10, ** p$<$0.05, *** p$<$0.01. $\widehat{\mathbf{W}}$ denotes the data-driven estimated spatial network matrix obtained using the BOLMT approach of \citet{JuodisKapetaniosSarafidis2025}. $\mathbf{W}_{1}$ denotes the contiguity matrix; $\mathbf{W}_{2}$ and $\mathbf{W}_{3}$ denote inverse-distance matrices with 10th and 5th percentile cutoffs, respectively; $\mathbf{W}_{4}$ and $\mathbf{W}_{5}$ denote trade-based weighting matrices, with $\mathbf{W}_{4}$ time-averaged and $\mathbf{W}_{5}$ time-varying.
}
\end{table}

\begin{figure}[H]
    \centering
    \makebox[\textwidth][c]{%
        \includegraphics[scale=1.0]{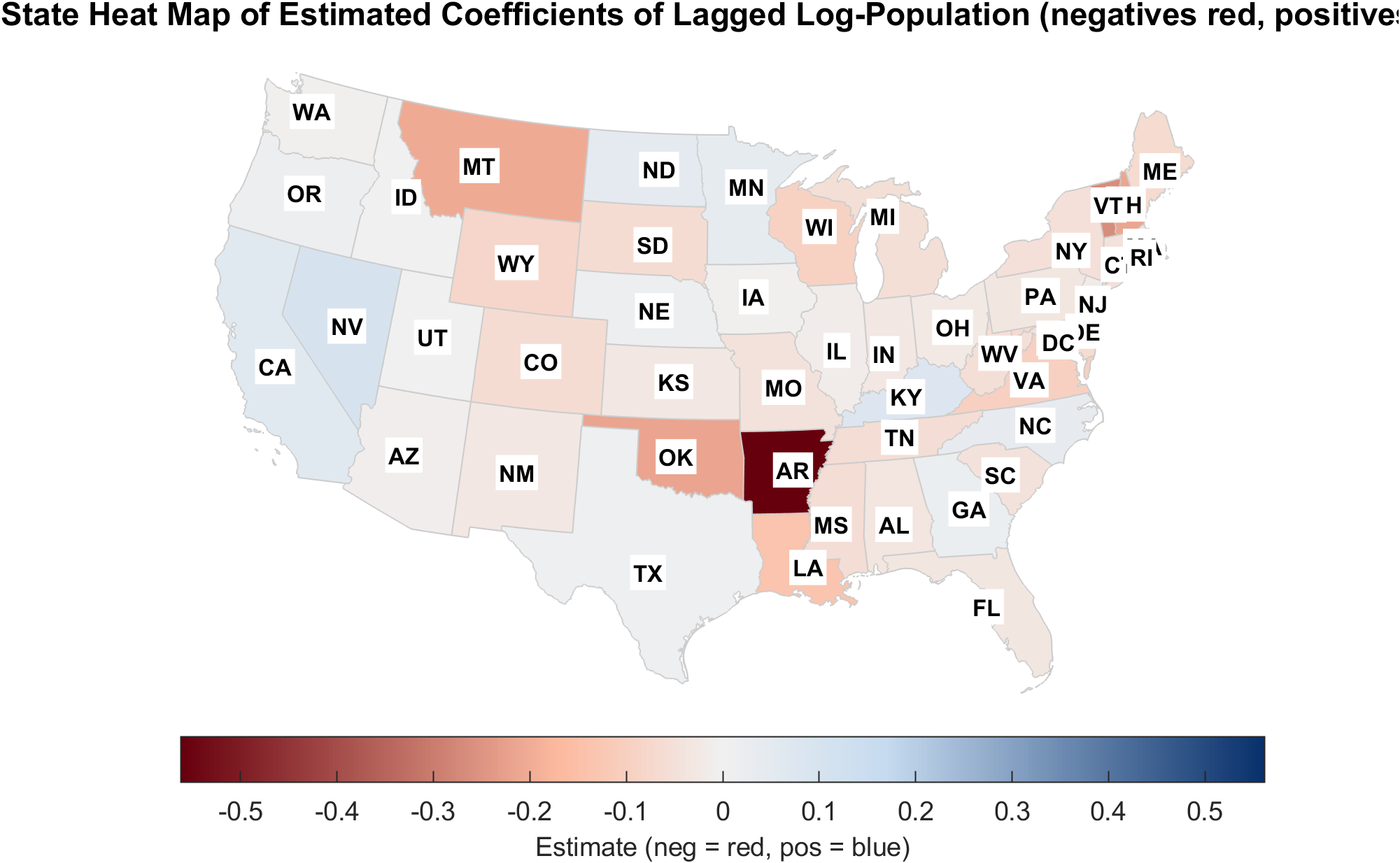}%
    }
    \caption{Heat Map of Estimated Coefficients of Lagged Log-Population}
    \label{fig:heatmap_population}
\end{figure}

Figure~\ref{fig:heatmap_population} reports a heatmap of the estimated coefficients $\widehat{\delta}_{i}$ across U.S. states. These coefficients measure the sensitivity of population growth to the lagged log level of population and thus govern transitional dynamics. By construction, $\delta_i<0$ implies conditional convergence: states with larger initial populations grow more slowly. On the other hand, $\delta_i>0$ indicates divergence, with more populous states expanding faster.

The graph reveals non-negligible heterogeneity in both the sign and magnitude of $\delta_i$. Most states ($73.5\%$) display negative coefficients (shades of red), signalling that population growth in the United States over the sample period was broadly mean-reverting. However, the degree of convergence varies sharply across space. States such as Arkansas, Oklahoma, Vermont, Montana, New Hampshire, and Louisiana exhibit strongly negative coefficients (dark red), suggesting pronounced transitional dynamics and a high speed of adjustment toward steady-state population levels. By contrast, much of the Midwest and Southeast is coloured pale red, indicating only modest convergence. A second, smaller cluster of states, including California, Nevada, Minnesota, and North Carolina, shows slightly positive coefficients (light blue), implying localized divergence. These states are characterised by persistently high in-migration, dynamic labour markets, and strong productivity or amenity advantages, which can sustain cumulative population growth over time (see, e.g., \citet{EdwardsGrobar2002}). 



To shed light on the economic mechanisms potentially underlying this heterogeneity, we contrast trade exposure between converging and diverging states. States with positive values of $\widehat{\delta}_i$, indicating divergence, display substantially higher average trade volumes than states with negative $\widehat{\delta}_i$, which are characterised by convergence. The difference in means is economically large and statistically meaningful (p-value = 0.044), suggesting that stronger trade exposure is associated with divergent population dynamics rather than mean reversion. This association is consistent with the idea that stronger trade linkages support continued economic activity and population growth, leading to divergent population dynamics rather than convergence. By contrast, states with weaker trade exposure tend to exhibit slower growth in larger populations and more pronounced convergence dynamics.\footnote{Interestingly, the same pattern does not emerge when states are grouped by average in-migration flows, amenity measures, or total factor productivity. In these cases, differences between converging and diverging states are not statistically distinguishable, suggesting that divergence in population dynamics is not simply driven by higher inflows, amenities, or productivity levels, but is more closely associated with differences in trade exposure.}


\subsection{Marginal Effects}

Table~\ref{tab:me_2siv} reports the estimated average marginal effects of the determinants of state population growth under the homogeneous specification, based on the spatial 2SIV framework of \citet{CuiSarafidisYamagata2023}. Three types of effects are shown: direct, indirect (spatial spillovers), and total. Overall, the estimated coefficients are highly stable across specifications in both sign and magnitude, indicating that the results are robust to the choice of spatial weighting matrix.

For the endogenously estimated matrix $\widehat{\mathbf{W}}$, all key covariates exhibit statistically significant direct effects, with expected signs. Total factor productivity (TFP), amenities, and labour income are all positively associated with population growth, while migration costs exert a negative effect. The relative magnitude of the direct/indirect effects indicate that roughly $35.3\%$ of the total effects on population growth operates through spatial spillovers. This reflects a substantial spatial diffusion component in the model.\footnote{Note that in this model, the ratio of indirect to total effects is, by construction, identical across all covariates. The slight differences observed here arise solely from rounding of the reported coefficients.}
The total effects, computed as the sum of direct and indirect components, are uniformly positive and statistically significant for amenities, and labour income across all weighting matrices, while migration costs remain consistently negative. 

Table~\ref{tab:me_mgiv} presents the corresponding average marginal effects from the heterogeneous specification, estimated using the spatial MGIV estimator of \citet{ChenCuiSarafidisYamagata2025}. The direct effects of labour income and amenities remain strongly significant across all spatial weighting matrices, while migration costs continue to exhibit a robust negative impact. By contrast, total factor productivity is statistically significant only when the weighting matrix is estimated from the data, $\widehat{\mathbf{W}}$, and becomes insignificant under all alternative matrices, highlighting the empirical relevance of endogenously determining the spatial structure. Relative to the homogeneous (pooled 2SIV) specification, the share of total marginal effects attributable to direct effects is very similar, at approximately 65.6\%. Nonetheless, the estimated direct (and total) effects under MGIV are generally larger in magnitude, suggesting that pooled estimators understate the true responsiveness of population growth to its determinants when slope heterogeneity is ignored. As emphasised by \citet{BreitungSalish2021}, when slope coefficients vary systematically with regressors, estimators that impose common slopes recover pseudo parameters that need not coincide with population-average responses. By contrast, MGIV constructs average marginal effects from state-specific responses, yielding both larger magnitudes and an economically meaningful population-level interpretation. On balance, the larger estimated effects under MGIV indicate that accounting for slope heterogeneity yields a more flexible and empirically credible representation of population growth dynamics.

\begin{table}[H]
\centering
\caption{Average Marginal Effects for State Population Growth (2SIV)}
\label{tab:me_2siv}
\small
\begin{tabular}{lcccccc}
\toprule
 & $\widehat{\mathbf{W}}$ & $\mathbf{W}_{1}$ & $\mathbf{W}_{2}$ & $\mathbf{W}_{3}$ & $\mathbf{W}_{4}$ & $\mathbf{W}_{5}$ \\
\midrule
\multicolumn{7}{l}{\textbf{Direct effects}}\\
\addlinespace[2pt]
Lagged population & -0.016*** & -0.023*** & -0.014*** & -0.019*** & -0.025*** & -0.026 \\
 & (0.005) & (0.005) & (0.004) & (0.005) & (0.005) & $-$ \\
TFP & 0.000 & 0.002 & 0.002** & 0.003** & 0.002** & 0.001 \\
 & (0.001) & (0.001) & (0.001) & (0.001) & (0.001) & $-$ \\
Amenities & 0.005*** & 0.006*** & 0.006*** & 0.006*** & 0.007*** & 0.006 \\
 & (0.000) & (0.000) & (0.000) & (0.000) & (0.000) & $-$ \\
Labour income & 0.038*** & 0.032*** & 0.033*** & 0.032*** & 0.032*** & 0.033 \\
 & (0.004) & (0.004) & (0.005) & (0.005) & (0.003) & $-$ \\
Migration cost & -0.002*** & -0.002*** & -0.002*** & -0.002*** & -0.002*** & -0.002 \\
 & (0.000) & (0.000) & (0.000) & (0.000) & (0.000) & $-$ \\
\addlinespace[4pt]
\multicolumn{7}{l}{\textbf{Indirect effects}}\\
\addlinespace[2pt]
Lagged population & -0.008*** & -0.014*** & -0.003*** & -0.003*** & -0.008 & -0.029 \\
 & (0.002) & (0.004) & (0.001) & (0.001) & (0.005) & $-$ \\
TFP & 0.000 & 0.001 & 0.000 & 0.000 & 0.000 & 0.001 \\
 & (0.001) & (0.001) & (0.000) & (0.000) & (0.000) & $-$ \\
Amenities & 0.003*** & 0.004*** & 0.001*** & 0.001*** & 0.002 & 0.007 \\
 & (0.000) & (0.001) & (0.000) & (0.000) & (0.001) & $-$ \\
Labour income & 0.019*** & 0.018*** & 0.006*** & 0.005*** & 0.010 & 0.038 \\
 & (0.005) & (0.005) & (0.002) & (0.001) & (0.007) & $-$ \\
Migration cost & -0.001*** & -0.001*** & -0.000*** & -0.000*** & -0.001 & -0.002 \\
 & (0.000) & (0.000) & (0.000) & (0.000) & (0.000) & $-$ \\
\addlinespace[4pt]
\multicolumn{7}{l}{\textbf{Total effects}}\\
\addlinespace[2pt]
Lagged population & -0.024*** & -0.037*** & -0.017*** & -0.022*** & -0.033*** & -0.055 \\
 & (0.008) & (0.007) & (0.005) & (0.005) & (0.006) & $-$ \\
TFP & 0.000 & 0.003 & 0.003** & 0.003** & 0.002** & 0.002 \\
 & (0.002) & (0.002) & (0.001) & (0.001) & (0.001) & $-$ \\
Amenities & 0.008*** & 0.010*** & 0.007*** & 0.007*** & 0.009*** & 0.013 \\
 & (0.001) & (0.001) & (0.001) & (0.001) & (0.001) & $-$ \\
Labour income & 0.057*** & 0.050*** & 0.039*** & 0.036*** & 0.042*** & 0.071 \\
 & (0.008) & (0.008) & (0.006) & (0.005) & (0.008) & $-$ \\
Migration cost & -0.002*** & -0.003*** & -0.003*** & -0.003*** & -0.003*** & -0.004 \\
 & (0.000) & (0.000) & (0.000) & (0.000) & (0.001) & $-$ \\
\midrule
N & 49 & 49 & 49 & 49 & 49 & 49 \\
\bottomrule
\end{tabular}
\vspace{0.5em}
\parbox{\linewidth}{
\small
\textit{\textbf{Notes}:} Standard errors in parentheses. * p$<$0.10, ** p$<$0.05, *** p$<$0.01. Standard errors for time-varying marginal effects are not computed. $\widehat{\mathbf{W}}$ denotes the data-driven spatial network estimated using the approach of \citet{JuodisKapetaniosSarafidis2025}. $\mathbf{W}_{1}$ is the contiguity matrix; $\mathbf{W}_{2}$–$\mathbf{W}_{3}$ are inverse-distance matrices (10th and 5th percentile cutoffs); $\mathbf{W}_{4}$–$\mathbf{W}_{5}$ are trade-based matrices (time-averaged and time-varying).
}
\end{table}

\begin{table}[H]
\centering
\caption{Average Marginal Effects for State Population Growth (MGIV)}
\label{tab:me_mgiv}
\small
\begin{tabular}{lcccccc}
\toprule
 & $\widehat{\mathbf{W}}$ & $\mathbf{W}_{1}$ & $\mathbf{W}_{2}$ & $\mathbf{W}_{3}$ & $\mathbf{W}_{4}$ & $\mathbf{W}_{5}$ \\
\midrule
\multicolumn{7}{l}{\textbf{Direct effects}}\\
\addlinespace[2pt]
Lagged population & -0.051*** & -0.046*** & -0.048*** & -0.050*** & -0.050*** & -0.052 \\
 & (0.015) & (0.013) & (0.013) & (0.014) & (0.014) & $-$ \\
TFP & 0.005** & 0.001 & 0.003 & 0.003 & 0.005 & 0.004 \\
 & (0.003) & (0.003) & (0.003) & (0.003) & (0.004) & $-$ \\
Amenities & 0.007*** & 0.006*** & 0.006*** & 0.006*** & 0.006*** & 0.006 \\
 & (0.001) & (0.001) & (0.001) & (0.001) & (0.001) & $-$ \\
Labour income & 0.051*** & 0.051*** & 0.052*** & 0.053*** & 0.052*** & 0.051 \\
 & (0.012) & (0.011) & (0.012) & (0.013) & (0.011) & $-$ \\
Migration cost & -0.004*** & -0.002*** & -0.003*** & -0.002*** & -0.003*** & -0.003 \\
 & (0.001) & (0.001) & (0.001) & (0.001) & (0.001) & $-$ \\
\addlinespace[4pt]
\multicolumn{7}{l}{\textbf{Indirect effects}}\\
\addlinespace[2pt]
Lagged population & -0.028** & -0.049*** & -0.039*** & -0.030*** & -0.127 & -0.075 \\
 & (0.011) & (0.018) & (0.013) & (0.009) & (0.135) & $-$ \\
TFP & 0.003** & 0.001 & 0.002 & 0.002 & 0.014 & 0.005 \\
 & (0.001) & (0.003) & (0.003) & (0.002) & (0.016) & $-$ \\
Amenities & 0.004*** & 0.006*** & 0.005*** & 0.004*** & 0.016 & 0.008 \\
 & (0.001) & (0.002) & (0.002) & (0.001) & (0.017) & $-$ \\
Labour income & 0.028*** & 0.054** & 0.043*** & 0.032*** & 0.131 & 0.074 \\
 & (0.010) & (0.022) & (0.014) & (0.009) & (0.137) & $-$ \\
Migration cost & -0.002*** & -0.002*** & -0.002*** & -0.001*** & -0.008 & -0.004 \\
 & (0.000) & (0.001) & (0.001) & (0.000) & (0.008) & $-$ \\
\addlinespace[4pt]
\multicolumn{7}{l}{\textbf{Total effects}}\\
\addlinespace[2pt]
Lagged population & -0.079*** & -0.095*** & -0.087*** & -0.080*** & -0.176 & -0.128 \\
 & (0.025) & (0.028) & (0.025) & (0.023) & (0.140) & $-$ \\
 TFP & 0.008** & 0.003 & 0.005 & 0.004 & 0.019 & 0.009 \\
 & (0.004) & (0.007) & (0.006) & (0.005) & (0.018) & $-$ \\
Amenities & 0.011*** & 0.011*** & 0.011*** & 0.010*** & 0.022 & 0.014 \\
 & (0.003) & (0.003) & (0.003) & (0.002) & (0.017) & $-$ \\
Labour income & 0.078*** & 0.105*** & 0.095*** & 0.085*** & 0.183 & 0.126 \\
 & (0.020) & (0.031) & (0.025) & (0.021) & (0.140) & $-$ \\
Migration cost & -0.005*** & -0.005*** & -0.005*** & -0.004*** & -0.010 & -0.007 \\
 & (0.001) & (0.001) & (0.001) & (0.001) & (0.008) & $-$ \\
\midrule
$N$ & 49 & 49 & 49 & 49 & 49 & 49 \\
\bottomrule
\end{tabular}
\vspace{0.5em}
\parbox{\linewidth}{
\small
\textit{\textbf{Notes}:} Standard errors in parentheses. * p$<$0.10, ** p$<$0.05, *** p$<$0.01. Standard errors for time-varying marginal effects are not computed. $\widehat{\mathbf{W}}$ denotes the data-driven spatial network estimated using the approach of \citet{JuodisKapetaniosSarafidis2025}. $\mathbf{W}_{1}$ is the contiguity matrix; $\mathbf{W}_{2}$–$\mathbf{W}_{3}$ are inverse-distance matrices (10th and 5th percentile cutoffs); $\mathbf{W}_{4}$–$\mathbf{W}_{5}$ are trade-based matrices (time-averaged and time-varying).
}
\end{table}

\begin{figure}[H]
    \centering
    \includegraphics[width=0.8\textwidth]{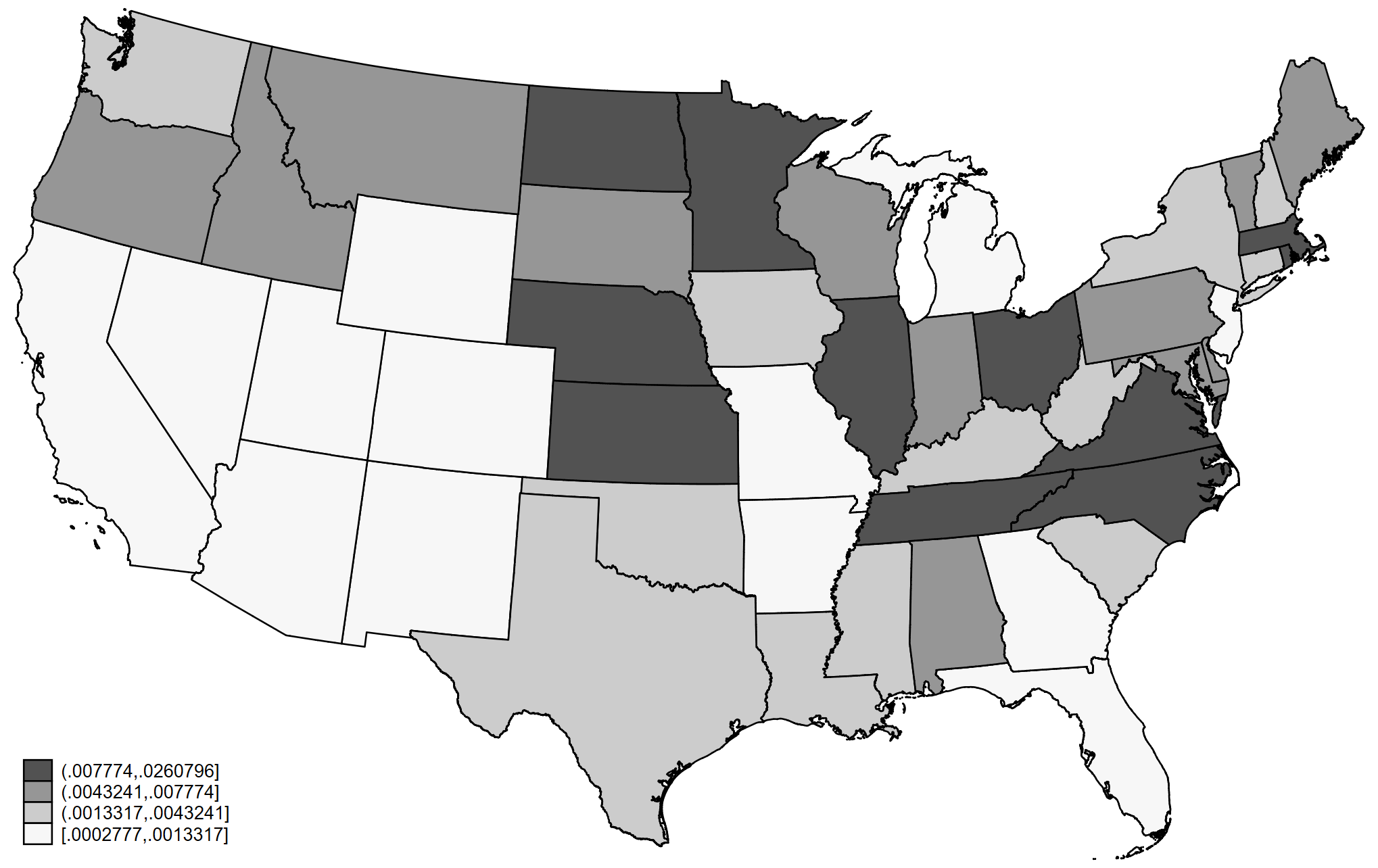}
    \caption{Population Spillovers Originating from States in the Midwest Region}
    \label{fig:impacts_midwest}
\end{figure}

Figure~\ref{fig:impacts_midwest} illustrates indirect (spillover) effects in population growth originating from states in the Midwest, obtained using the MGIV estimator with $\widehat{\mathbf{W}}$. The darker shading indicates that states are subject to stronger spillovers due to growth originating in the Midwest. In contrast, lighter shades indicate weaker effects.

The results reveal a clear pattern of spatial diffusion in which the Midwest exerts non-negligible influences beyond its own borders. The strongest spillovers arise in neighbouring states (particularly in the South Atlantic, East South Central, and West North Central divisions) in a manner consistent with both geographic proximity and economic linkages\footnote{For example, labour mobility, trade in goods and services, and commuting flows between the industrial Midwest and adjoining regions.}. However, diffusion does not uniformly follow a simple distance-decay pattern: several non-adjacent states, such as Washington and Oregon, also register relatively strong indirect effects. Pronounced spillovers are likewise observed in Utah and Arizona, both of which are economically dynamic and deeply integrated into national production and logistics networks. These states have experienced sustained industrial expansion and population inflows partly linked to relocations and trade flows originating in the Midwest.

Collectively, these findings indicate that population dynamics propagate through broader economic and migration networks that extend beyond immediate contiguity. This evidence reinforces the use of an endogenously estimated spatial weighting matrix, $\widehat{\mathbf{W}}$, which allows economically meaningful, non-local connections to emerge from the data rather than being imposed \emph{a priori} via contiguity- or distance-based rules.


\subsection{Network Mapping}

A central innovation of this paper is the use of a data-driven approach to estimate the
spatial weighting matrix that captures inter-state interactions. Rather than relying exclusively on geographic proximity, our method uncovers underlying network connections through which population growth propagates. 

\begin{figure}[H]
    \centering
    \makebox[\textwidth][c]{%
        \includegraphics[scale=1.0]{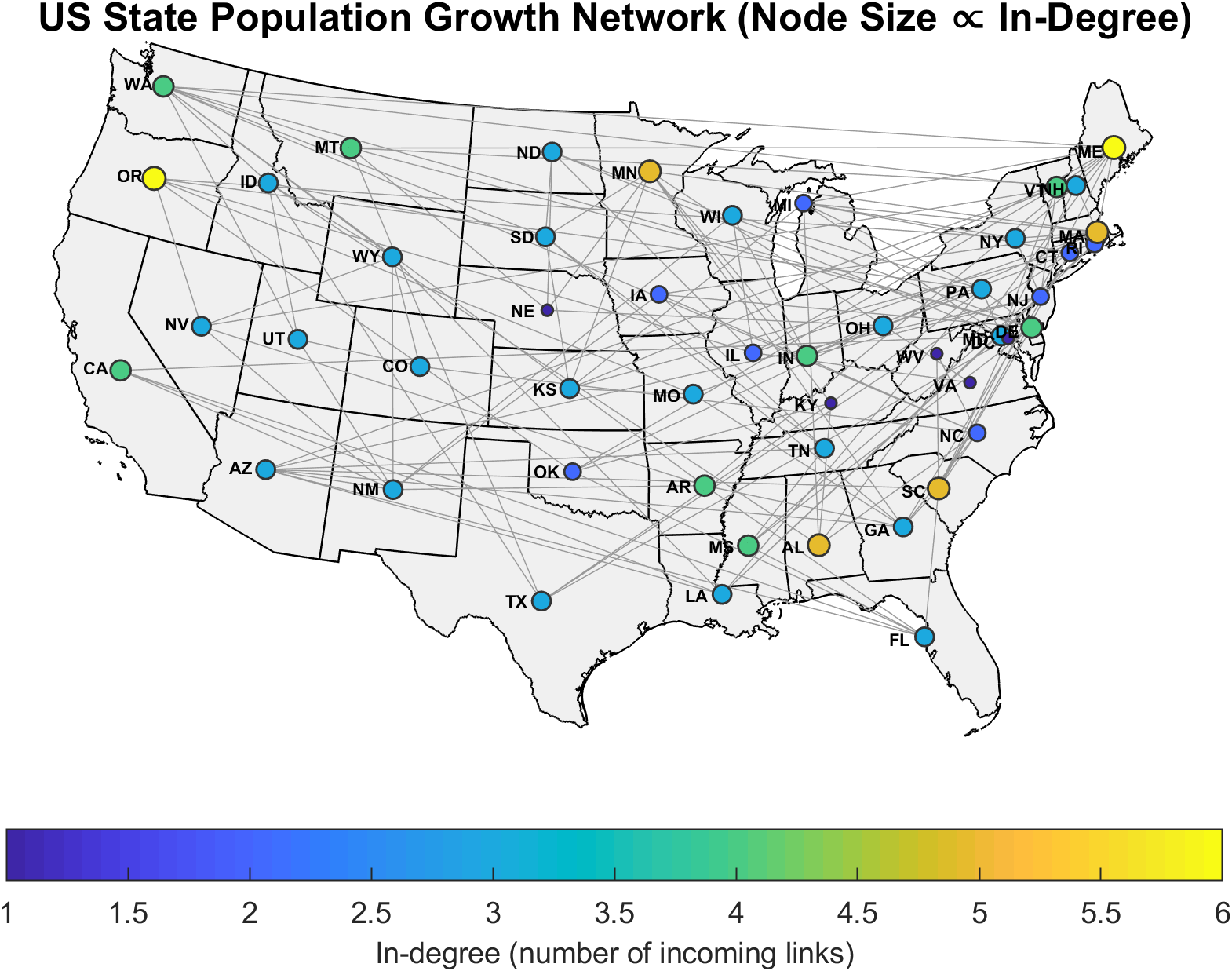}%
    }
    \caption{U.S. States Population Growth Network}
    \label{fig:US_State_Population_Network}
\end{figure}

Figure \ref{fig:US_State_Population_Network} below visualises the estimated spatial weighting matrix $\widehat{\mathbf{W}}$ recovered using the BOLMT procedure. Each node represents a U.S. state, and a directed edge from state $j$ to state $i$ is drawn whenever the corresponding element $\widehat{w}_{i,j}$ is non-zero, indicating a statistically selected influence of state $j$ on state $i$’s population growth. Node colour and size encode in-degree; i.e., the number of other states exerting a statistically significant influence on a given state’s population growth.

Although the estimated network is sparse overall (link density 0.66\%), it displays pronounced heterogeneity in connectivity. Importantly, the presence of long-distance connections indicates that population spillovers are not confined to geographic proximity.

States with larger, warmer-coloured nodes (prominently in the Northeast, such as Massachusetts, New York, and Maine, and in parts of the Pacific Northwest like Oregon and Washington), exhibit relatively high in-degree values, consistent with population dynamics shaped by multiple external influences and dense inter-state integration. 
By contrast, smaller, cooler-coloured nodes (for instance in parts of the Midwest, such as Nebraska and North Dakota), display lower in-degree, indicating that their growth patterns are less dependent on spillovers from surrounding states and more self-contained.

Complementing this in-degree perspective, the distribution of out-degree centrality (the number of directed links a state sends to others) shows limited dispersion rather than heavy-tailed behaviour. In particular, there is no clear evidence of hubs or dominant states that broadcast disproportionate influence across the network. Instead, the structure is sparse and decentralised, with influence more evenly distributed and spillovers propagating through indirect and cumulative linkages rather than being driven by a small set of key states. 

\subsection{Network Formation and Homophily Sources}

The spatial weighting matrix $\widehat{\mathbf{W}}$ summarises the estimated pattern of connectivity across states, but it does not directly inform on the forces shaping these connections. To shed light on potential mechanisms underlying the formation of inter-state links, the analysis proceeds in two steps. We first assess regional homophily by examining whether links are disproportionately concentrated within the same U.S. Census divisions. We then examine whether the intensity of bilateral trade and migration inflows from state $j$ to state $i$ is associated with a higher likelihood that the two states are linked in the estimated network.


\subsubsection{Regional Homophily}
To assess the extent to which \textit{geographic contiguity} influences the formation of links, we adopt the following empirical setup.
Let $s_i\in\{1,\dots,9\}$ index state $i$’s U.S. Census Bureau division.\footnote{Codes:
1 = East South Central; 2 = Mountain; 3 = West South Central; 4 = Pacific; 5 = New England; 6 = South Atlantic (includes D.C.); 7 = East North Central; 8 = West North Central; 9 = Middle Atlantic.}

We first construct a binary indicator matrix capturing whether any two states belong to the same division:
\begin{equation}
s_{i,j} = 
\begin{cases}
1 & \text{if } s_i = s_j \text{ and } i \neq j, \\
0 & \text{otherwise}.
\end{cases}
\end{equation}

The total number of same-division links observed in the estimated network $\widehat{\mathbf{W}}$ is then computed as:
\begin{equation}
L_{\text{same}} = \sum_{i \ne j} \widehat{w}_{i,j} \cdot s_{i,j},
\end{equation}
where $\widehat{w}_{i,j}$ denotes the $(i,j)$th entry of $\widehat{\mathbf{W}}$.

To assess whether the observed concentration of links within divisions exceeds chance, we conduct a permutation test following \citet{HannemanRiddle2005} and \citet{LaFondNeville2010}. Under the null of no division-based clustering, division labels are unrelated to the network structure. We therefore permute the division labels while preserving division sizes (i.e., we randomly reassign the labels $\{s_i\}_{i=1}^N$ among states):

\begin{itemize}
    \item For each permutation $b=1,\dots,B$, draw a random permutation of the labels to obtain $\{s_i^{(b)}\}_{i=1}^N$.
    \item Using $\{s_i^{(b)}\}$, recompute the same-division indicator matrix $s_{i,j}^{(b)}=\mathbf{1}\{s_i^{(b)}=s_j^{(b)},\, i\neq j\}$.
    \item Compute the permuted within-division link weight:
    \[
      L_{\text{same}}^{(b)} \;=\; \sum_{i\neq j} \widehat w_{i,j}\, s_{i,j}^{(b)}.
    \]
    \item Estimate the one-sided $p$-value (greater mass on within-division links than by chance) as
    \[
      p \;=\; \frac{\sum_{b=1}^{B} \mathbf{1}\!\left\{ L_{\text{same}}^{(b)} \,\ge\, L_{\text{same}} \right\}}{B}.
    \]
\end{itemize}
We set $B = 10{,}000$. A small $p$-value indicates that the observed weight on same-division links is significantly larger than would be expected under random assignment, providing evidence of homophily.

The observed number of same-division links is $L_{\text{same}}=37$, and the permutation test yields $p<0.001$, providing statistically significant evidence of homophily at the 5\% level. Given the network's high sparsity (link density 0.66\%), even moderate within-division clustering is unlikely to arise by chance. These findings indicate that geographic proximity contributes meaningfully to network formation.

To quantify not only the presence but also the magnitude of homophily, we compute the proportion of links that occur between states in the same region:
\begin{equation}
\widehat h \;=\; \frac{L_{\text{same}}}{L_{\text{total}}}; 
\qquad
L_{\text{same}} \;=\; \sum_{i\ne j} \widehat w_{ij} \, s_{ij};
\qquad
L_{\text{total}} \;=\; \sum_{i\ne j} \widehat w_{ij}.
\end{equation}

\noindent
Thus, $\widehat h$ is the share of total \emph{link weight} assigned to same-division pairs.
The expected proportion under the null hypothesis is given by:
\begin{equation}
\mathbb{E}[\widehat{h}_{\text{null}}] = \frac{1}{B} \sum_{b=1}^{B} \frac{L_{\text{same}}^{(b)}}{L_{\text{total}}}.
\end{equation}

We report two complementary measures:
\begin{align}
\text{Relative Homophily Index} &= \frac{\widehat{h}}{\mathbb{E}[\widehat{h}_{\text{null}}]} = \frac{0.245}{0.1085} = 2.259, \\
\text{Homophily Excess (in p.p.)} &= \widehat{h} - \mathbb{E}[\widehat{h}_{\text{null}}] = 0.137.
\end{align}

The Relative Homophily Index of $2.259$ indicates that links between same-division states occur about 2.26 times as often (that is, approximately 126\% more frequently) as would be expected under random assignment. The Homophily Excess expresses this in absolute terms: 13.7 percentage points more of the observed link weight is within-division than under the null. Together, these measures provide a clear and interpretable assessment of homophily in the network, capturing both its direction (positive homophily) and magnitude (relative and absolute strength).

\subsubsection{Homophily in Bilateral Trade and Migration Inflows}

We next examine whether inter-state link formation is systematically related to bilateral trade and migration inflows. Since these variables are continuous rather than categorical, a different empirical strategy is required. Let $d_{1,i,j,t}$ and $d_{2,i,j,t}$ denote the logged values of migration and trade inflows from state $j$ to state $i$ at time $t$, respectively, and define the corresponding time averages as
\begin{equation*}
\overline{d}_{\ell,i,j} = \frac{1}{T} \sum_{t=1}^{T} d_{\ell,i,j,t}; \qquad \ell = 1,2.
\end{equation*}
For each variable $\ell$, we collect these averages into the $N \times N$ matrix $\mathbf{P}_{\ell} = [\overline{d}_{\ell,i,j}]_{i,j=1}^{N}$.

To relate these bilateral measures to the estimated network, we vectorise the off-diagonal elements of both the estimated network matrix and $\mathbf{P}_{\ell}$. Specifically, let
\[
\widetilde{\mathbf{w}} = \operatorname{vec}_{\text{off}}(\widetilde{\mathbf{W}}),
\qquad
\mathbf{p}_{\ell} = \operatorname{vec}_{\text{off}}(\mathbf{P}_{\ell}),
\]
where $\widetilde{\mathbf{W}}$ is obtained from $\widehat{\mathbf{W}}$ by replacing non-zero entries with ones, and $\operatorname{vec}_{\text{off}}(\cdot)$ stacks all $i \neq j$ elements of a matrix into a column vector. Indexing the resulting off-diagonal pairs by $m = 1,\dots,M$, with $M = N(N-1)$, let $\widetilde{w}_{m}$ denote the $m$th element of $\widetilde{\mathbf{w}}$ and $p_{\ell,m}$ the corresponding element of $\mathbf{p}_{\ell}$.

We model link formation using the logistic regression
\begin{equation}
\label{eq:logit_homophily}
\log\!\left(\frac{\Pr(\widetilde{w}_{m}=1 \mid \mathbf{p}_{m})}{1 - \Pr(\widetilde{w}_{m}=1 \mid \mathbf{p}_{m})}\right)
= \alpha + \sum_{\ell=1}^{2} \pi_{\ell}\, p_{\ell,m},
\end{equation}
where $\mathbf{p}_{m} = (p_{1,m}, p_{2,m})^{\prime}$ collects the bilateral migration and trade measures, and $\boldsymbol{\pi} = (\pi_{1}, \pi_{2})^{\prime}$ is the corresponding coefficient vector. The parameters $\alpha$ and $\boldsymbol{\pi}$ are estimated using a bias-corrected logistic estimator.\footnote{This estimator is implemented via the user-written \texttt{relogit} command in Stata, which applies the rare-events logistic correction of \citet{King2001RareEvents}.} Bias correction is important in this setting because the estimated network is sparse, with a density of approximately $0.66\%$, implying that conventional maximum likelihood estimates would otherwise be subject to non-negligible small-sample bias.

Among the regressors considered, only logged bilateral migration inflows from state $j$ to state $i$ are statistically significant at the 1\% level. The estimated coefficient is $\widehat{\pi}_{1} = 0.141$ (p-value = 0.009), corresponding to an odds ratio of $\exp(0.141) \approx 1.15$. This implies that a one-unit increase in logged bilateral migration inflows from state $j$ to state $i$ is associated with approximately a 15\% increase in the odds that a directed link from $j$ to $i$ is selected. In other words, stronger bilateral migration flows are associated with a substantially higher likelihood that the two states are linked in the estimated network. This pattern suggests that the data-driven construction of the network recovers economically meaningful inter-state connections, rather than spurious correlations.


\section{Concluding Remarks}

Our analysis brings together endogenous network recovery, heterogeneous slopes, and interactive fixed effects within a single estimation framework to study the drivers and diffusion of U.S. state population growth. Empirically, we find broad yet uneven conditional convergence, stable effects of amenities, labour income, and migration frictions across alternative network specifications, and a productivity effect that becomes identifiable only when the spatial network is estimated from the data. Spatial spillovers are economically meaningful: indirect effects account for roughly one-third of total marginal impacts, and diffusion often extends beyond contiguous neighbours. The inferred network is sparse, with localized clusters of higher connectivity and clear regional homophily. The common factor error structure accounts for a  for at least 60\% of the residual variance. These features together highlight the value of modelling both network linkages and pervasive cross-sectional dependence.

The new spill-in/spill-out diagnostics open the door to targeted policy evaluation. For instance, mapping which regions transmit or absorb growth most strongly can inform where interventions are likely to propagate, and where complementary measures (housing supply, infrastructure, workforce mobility) are needed to translate local gains into broader regional outcomes.

\bibliographystyle{apalike}
\bibliography{bibliography}

\section*{Appendix A: Additional Information}

\renewcommand{\thetable}{A.\arabic{table}}
\setcounter{table}{0}

\clearpage
\begin{landscape}
\begin{table}[htbp]
\centering
\caption{Summary of Spatial Weighting Matrices}
\label{tab:weights}
\begin{tabular}{llp{7.2cm}p{3.8cm}}
\hline
\textbf{Label} & \textbf{Definition of $w_{i,j}$} & \textbf{Notes} & \textbf{Characteristics} \\
\hline

$\widehat{\mathbf{W}}$ &
Data-driven estimation. &
Estimated using the method of \citet{JuodisKapetaniosSarafidis2025}. Represents the inferred network of interactions among states. &
Density = 0.66\%; Avg. links per state = 3.08 \\
 
$\mathbf{W}_{1}$ &
$w_{i,j} =
\begin{cases}
1 & \text{if} \; j \; \text{shares a border with } i ,\\
0                  & \text{otherwise.}
\end{cases}$ &
Spatial contiguity matrix &
Symmetric; Density = 9.27\%; Avg. links per state = 4.45.\\
 
$\mathbf{W}_{2}$ &
$w_{i,j} =
\begin{cases}
\dfrac{1}{d_{i,j}} & \text{if } d_{i,j} \le c \text{ miles},\\
0                  & \text{otherwise.}
\end{cases}$ &
$d_{i,j}$ is the great-circle distance from the Haversine formula between the capital cities in states $i,j$. The threshold, c, is set equal to the 10th percentile of the distribution of $d_{i,j}$. &
Density = 10\%; Avg. links per state = 4.86. \\

$\mathbf{W}_{3}$&
$w_{i,j} =
\begin{cases}
\dfrac{1}{d_{i,j}} & \text{if } d_{i,j} \le c \text{ miles},\\
0                  & \text{otherwise.}
\end{cases}$ &
$d_{i,j}$ is as above. The threshold c is set equal to the 5th percentile of the distribution of $d_{i,j}$. &
Density = 5\%; Avg. links per state = 2.61. \\

$\mathbf{W}_{4}$ &
$w_{i,j}=d_{i,j}$ & $d_{i,j}$ is the average value of trade inflows from $j$ to $i$. $w_{i,j}\neq 0$ for $i \neq j$. &
Density = 50.89\%; Avg. links per state = 23.41. \\

$\mathbf{W}_{5}$ &
$w_{i,j,t}=d_{i,j,t}$ & $d_{i,j,t}$ is the value of trade inflows from state $j$ to $i$ at time $t$. $w_{i,j,t}\neq 0$ for $i \neq j$. &
Density = 52.32\%; Avg. links per state = 24.44. \\
 
\hline
\end{tabular}
\vspace{0.5em}
\begin{minipage}{1\linewidth}
\footnotesize \textbf{Note:} All weighting matrices are row-normalised so that each row sums to one. For $\mathbf{W}_{4}$ -$\mathbf{W}_{5}$, the cut-off used to calculate summary statistics is set at $w_{i,j} > 0.01$.
\end{minipage}
\end{table}
\end{landscape}

\end{document}